\documentclass[twocolumn]{aastex62}

\usepackage{newtxmath}
\submitjournal{ApJ Lett.}

\shorttitle{Circular polarized filaments}
\shortauthors{Sinha et al.}
\begin{document}

\title{Magnetized Current Filaments as a Source of Circularly Polarized Light}

\correspondingauthor{L. O. Silva}
\email{luis.silva@tecnico.ulisboa.pt}

\author{U. Sinha}
\author{J. Martins}
\author{J. Vieira}
\author{K. M. Schoeffler}
\affil{GoLP/Instituto de Plasmas e Fus\~ao Nuclear,
Instituto Superior T\'ecnico,\\
Universidade de Lisboa, 1049-001 Lisboa, Portugal\\}
\author{R. A. Fonseca}
\affiliation{DCTI/ISCTE Instituto Universit\'ario de Lisboa, 1649-026 Lisboa, Portugal}
\affiliation{GoLP/Instituto de Plasmas e Fus\~ao Nuclear,
Instituto Superior T\'ecnico,\\
Universidade de Lisboa, 1049-001 Lisboa, Portugal\\}
\author{L. O. Silva}
\affil{GoLP/Instituto de Plasmas e Fus\~ao Nuclear,
Instituto Superior T\'ecnico,\\
Universidade de Lisboa, 1049-001 Lisboa, Portugal\\}



\begin{abstract}

We show that the Weibel or currente filamentation instability can lead to the emission of circularly polarized radiation. Using particle-in-cell (PIC) simulations and a radiation post-processing numerical algorithm, we demonstrate that the level of circular polarization increases with the initial plasma magnetization, saturating at $\sim 13\%$ when the magnetization, given by the ratio of magnetic energy density to the electron kinetic energy density, is larger than 0.05. Furthermore, we show that this effect requires an ion-electron mass ratio greater than unity. %also find that the finite circular polarization is observed for a magnetized plasma composed of higher ion-electron mass ratios.
These findings, which could also be tested in currently available laboratory conditions, show that the recent observation of circular polarization in gamma ray burst afterglows could be attributed to the presence of magnetized current filaments driven by the Weibel or the current filamentation instability.

\end{abstract}


%% We recommend that authors also use the natbib \citepp
%% and \citept commands to identify citations.  The citations are

\section{Introduction} 
\label{sec:intro}

Understanding the origin of polarization in the radiation from charged particles is of central importance in the study of many astrophysical objects like gamma-ray-bursts (GRBs)~\cite{wiersema2014,troja2017}, supernova remnants (SNRs)~\cite{milne1974,bandiera2016}, active galactic nuclei (AGN)~\cite{rodriguez2018} and pulsar wind nebulae (PWN)~\cite{linden2015}. The vast majority of theoretical models predict low degrees of linear polarization and no circular polarization~\cite{medvedev&loeb1999,gruzinov&waxman1999,matsumiya2003,sagiv2004,toma2008}. This is in contrast with recent observations that demonstrated the existence of circularly polarized radiation emission in the afterglow of the GRB 121024A~\cite{wiersema2014}. This emphasizes the importance of reinvestigating previously theoretical formulations regarding the origin of circular polarization in these scenarios, and, in particular, at the plasma scale~\cite{sagiv2004,sinha}. Here we show that circularly polarized radiation emission can be observed in the context of the Weibel/current filamentation instability (WI/CFI)~\cite{weibel1959}, and we explore the physical conditions under which this can occur.

It has been shown that the WI/CFI mediated collisionless shocks could be a possible mechanism to explain the power law distribution of charged particles and sub-equipartition magnetic fields believed to be present in GRBs~\cite{spitkovsky2008b,martins2009a,sironi2009,fiuza2012,stockem2014,huntington}, and the radiation obtained from charged particle motion in these magnetic fields is attributed to synchrotron process~\cite{hededal2005,spitkovskyspectra}. However, although the mechanism of radiation in such collisionless shocks was demonstrated, little attention has been paid to the polarization of the radiation emitted.

In this Letter, we show that %in magnetized plasmas, 
plasmas in the presence of an ambient magnetic field, %
composed of a light (e.g. electrons) and heavy (e.g. protons) species, the radiation emitted by the plasma particles due to their motion in fields generated due to WI/CFI is partially circularly polarized. The trajectories of  charged particles in the WI/CFI fields are studied using two-dimensional particle-in-cell (PIC) code OSIRIS~\cite{osiris,fonseca2013}. A  semi-analytical model is developed to describe such motion and estimate the radiation spectra and the degree of circular polarization for a fixed observer for various pitch angles that an electron makes with the WI/CFI magnetic fields in initially magnetized and unmagnetized plasmas. Furthermore, the radiation spectra and the degree of circular polarization are computed from particle trajectories extracted directly from PIC simulations and post-processing them using the radiation code jRad\cite{jrad}. %Sinha et al. to be submitted)~\footnote{U. Sinha, C. Keitel, N. Kumar, to be submitted.   
% The approach of tracking particle trajectories from PIC simulation to compute the polarized radiation is also employed here while the physical configuration, interaction geometry and underlying physics are different in both manuscripts.}
We show that an initial magnetization leads to transverse symmetry breaking in the pitch angle distribution of electrons confined in a current filament, producing a finite circular polarization in the emitted radiation. Of particular interest is the degree of circular polarization associated with different initial magnetizations.

\begin{figure}
	\centering
	\noindent\includegraphics[width=3.0in]{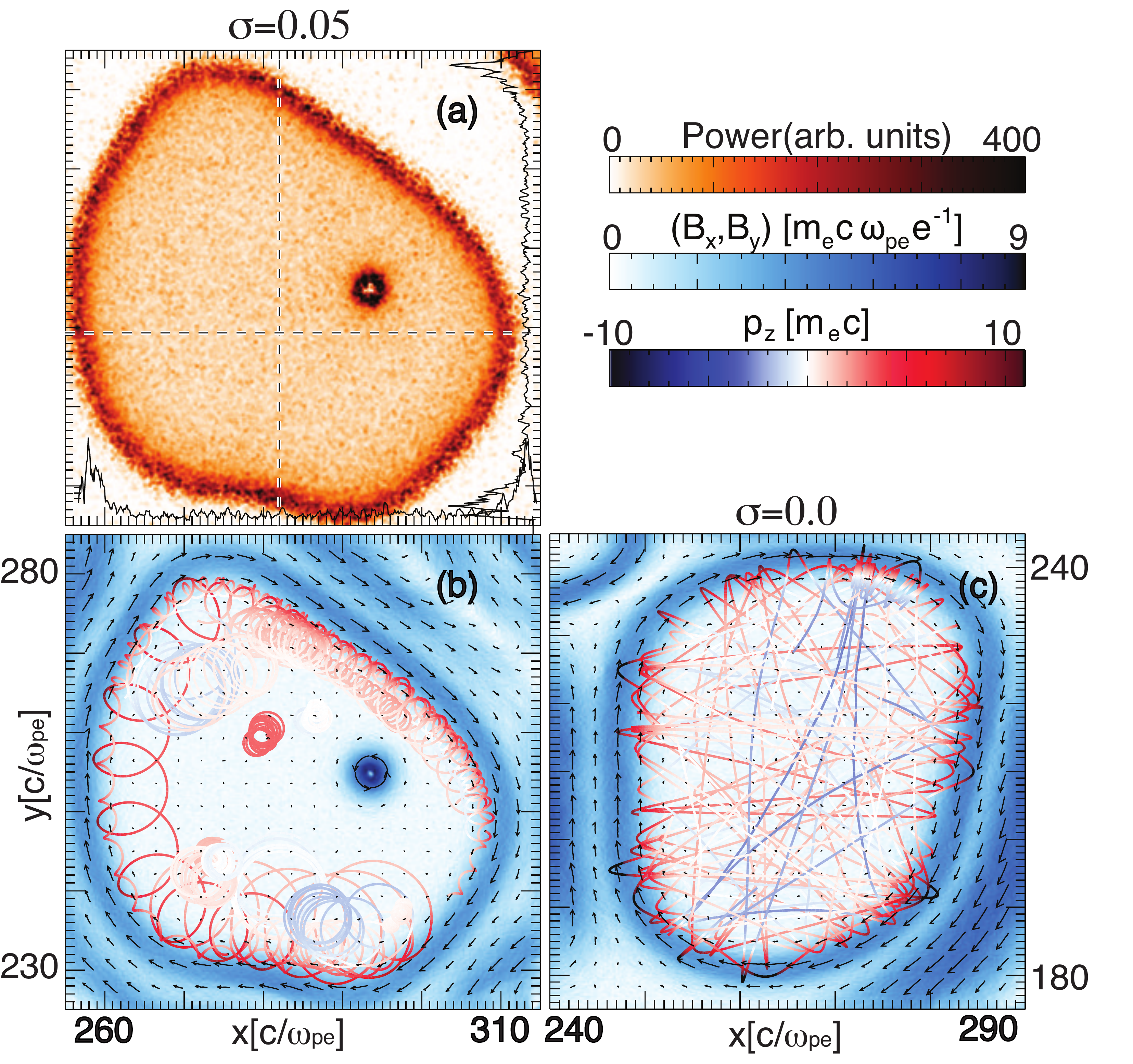}
	\caption{\label{fig1}(a) shows the spatial distribution of radiated power from electrons in a magnetized current filament of an electron-proton plasma ($m_i/m_e=1836$) with initial magnetization $\sigma=0.05$ and at time $t=9000\omega_{pe}^{-1}$. (b) shows the transverse magnetic field vectors (represented by arrows) arising due to WI/CFI for the same simulation as in (a). Trajectories of 50 electrons trapped in the current filament are shown from time $t_i$=8910$\omega_{pe}^{-1}$ to  $t_f$=9000$\omega_{pe}^{-1}$, with color scales representing their longitudinal momentum ($\textbf{p}_z$). The fields and tracks of electrons for the same time period in an unmagnetized electron-proton plasma are shown in (c).}
\end{figure}

We simulate a relativistic cold electron-proton plasma of uniform density and a relativistic factor $\gamma_0=20$ flowing through a background plasma at rest with identical %composition 
electron and proton densities. %and 
Both electrons and protons are initialized with a small isotropic thermal velocity $\beta_{th}\approx 10^{-3}$ (=$\sqrt{T/m c^2}$) to initiate the WI/CFI.
%a small temperature corresponding to a normalized thermal velocity $\beta_{th}\approx 10^{-3}$ in each direction.
The flow velocity (along z) is perpendicular to the simulation plane (see Fig.~\ref{fig1}). The computational domain is $512\times 512 (c/\omega_{pe})^2$ with periodic boundaries. The cell size is $\Delta x=\Delta y=0.1 c/\omega_{pe}$, time-step $\Delta t=0.045 \omega_{pe}^{-1}$ and each cell contains $3\times 3$ particles per cell per species, where $\omega_{pe}=(4\pi n_0e^2/m_e)^{1/2}$ is the plasma frequency, $n_0$ is the initial plasma density, $m_e$ is the electron mass, $e$ is the electron charge and $c$ is the velocity of light. We include a uniform out-of-plane (along z) initial magnetic field ($B_z$) measured in the frame of the background plasma corresponding to a magnetization of the flowing plasma $\sigma \equiv B_{z}^2/4\pi\gamma_0m_en_0c^2$ which we varied from $\sigma = 0.0-0.2$ for both signs of $B_z$. 
The WI/CFI in interpenetrating plasma flows give rise to current filaments with an associated azimuthal magnetic field ($B_\phi\hat{\textbf{e}}_\phi$) surrounding them. Because the electrons respond at time scales faster than the protons, a space charge radial electric field ($E_r\hat{\textbf{e}}_r$) develops at the edge of the electron filaments. The maximum electron WI/CFI growth rate is given by $\Gamma_{WI/CFI}\approx|\beta_0|/\sqrt{\gamma_0}\omega_{pe}$~\cite{silva2002,shukla2012}, where $\beta_0$ is the flow velocity normalized to c. For our $\gamma_0=20$ this yields $\Gamma\approx0.2\omega_{pe}$, and simulations show that the magnetic field saturates in a few 100$\omega_{pe}^{-1}$.

At the saturation of WI/CFI, there is a strong emission of radiation from the filament edges. This is illustrated in Fig.~\ref{fig1}a, which shows the radiated power ($P$) from the electrons in a single magnetized current filament. The radiated power is calculated using the relativistic generalization of Larmor's formula, $P=2/3(e^2/c)\gamma_e^6[\boldsymbol{\dot{\beta}}_e^2-(\boldsymbol{\beta}_e\times\boldsymbol{\dot{\beta}}_e)]$, where $\boldsymbol{\beta}_e$ is the electron velocity vector normalized to $c$, $\boldsymbol{\dot{\beta}}_e$ is the normalized electron acceleration and $\gamma_e$ is the electron Lorentz factor~\cite{jackson}. The peak value of $P$ at the filament edges is nearly ten times larger than that within the filament. This is because additional mutually perpendicular strong electric and magnetic fields resulting from WI/CFI leads to a relatively stronger $\boldsymbol{\dot{\beta}}_e$ perpendicular to $\boldsymbol{\beta}_e$ and increased kinetic energies of the electrons at the filament edges, whereas the electrons at the center gyrate only under the influence of the initial magnetic field and the longitudinal momentum ($p_z$) is much larger than the transverse momentum ($p_x,p_y$). Hence, the electrons at the filament edge primarily contribute to the radiated energy.
The electric field of the radiation emitted by an electron is given by~\cite{jackson},
\begin{equation}
\textbf{E}_\mathrm{rad}(\textbf{r},t) =\frac{e}{c} \Big[\frac{\textbf{n}\times[(\textbf{n}-\boldsymbol{\beta}_e)\times\boldsymbol{\dot{\beta}}_e]}{(1-\boldsymbol{\beta}_e\cdot\textbf{n})^3R}\Big]_{ret}\label{erad_xt}
\end{equation}
where $\textbf{n}$ is the unit vector from the position of the charge to the observer at a distance $R$ from the particle. The quantities are evaluated at the retarded time $t^\prime=t-R/c$. The spectrum of the radiated electric field is determined from the Fourier transform of Eq.\ref{erad_xt}~\cite{jackson},
%\begin{widetext}
\begin{eqnarray}
\textbf{E}_\mathrm{rad}(\omega)&=&\Big(\frac{e^2}{8\pi^2c}\Big)^{1/2}
\int_{-\infty}^{\infty}\frac{\textbf{n}\times[(\textbf{n}-\boldsymbol{\beta}_e)\times\boldsymbol{\dot{\beta}}_e]}{(1-\boldsymbol{\beta}_e\cdot\textbf{n})^2}\nonumber\\
&\times& e^{i\omega(t^\prime-\textbf{n}\cdot\textbf{r}(t^\prime)/c)}dt^\prime\label{erad_omega}
\end{eqnarray}
%\end{widetext}

The information about the instantaneous position, velocity and acceleration of the electrons can be obtained from the electron trajectories. The electron trajectories, superimposed on the magnetic fields due to WI/CFI for magnetized and unmagnetized plasmas, are shown in Figs.1b and 1c respectively. The electrons have enhanced longitudinal momentum ($p_z$) at the filament edge, which indicates that most of the radiation emitted from them will be collimated in the direction perpendicular to the simulation plane. An initial magnetization causes the electrons to drift along the azimuthal direction, due to a combination of $\textbf{E}\times\textbf{B}$ and $\nabla\textbf{B}$ drifts, executing a correlated motion (Fig.~\ref{fig1}b). On the other hand, in an unmagnetized plasma, the electrons scatter at all possible angles from the WI/CFI fields (Fig.~\ref{fig1}c).

To formulate a model describing the motion of charged particles in WI/CFI fields, we consider a cylindrical current filament with electrons flowing along the positive z-direction ($\hat{\textbf{e}}_z$). The electric and magnetic fields can be described as $\textbf{E}=E_rS(r)\hat{\text{e}}_r$ and $\textbf{B}=B_\phi S(r)\hat{\textbf{e}}_\phi+B_{z}\hat{\textbf{e}}_z$ respectively, where $E_r$ and $B_\phi$ are the respective amplitudes of the $\textbf{E}$ and $\textbf{B}$ fields in the radial and azimuthal directions, $B_z$ is the initial magnetic field and $S(r)$ is the spatial profile. At saturation, the spatial profile of the fields from the simulation closely resemble a Gaussian with FWHM$\sim$2.5$c/\omega_{pe}$. Hence, we substitute $S(r)=\exp[-(r-r_0)^2/2\eta^2]$ where $r_0$ is the filament radius and $\eta=1/\sqrt{2}$. The equation for the electron trajectory obtained by combining the momentum and energy equation is,
\begin{equation}
	\frac{d\boldsymbol{\beta}_e}{dt}=-\frac{1}{\gamma_e}[\textbf{E}+\boldsymbol{\beta}_e\times \textbf{B}-\boldsymbol{\beta}_e(\textbf{E}\cdot\boldsymbol{\beta}_e)]\label{track_sa}
\end{equation}
where the normalizations $\textbf{E}\equiv e\textbf{E}/(m_e\omega_{pe}c)$; $\textbf{B}\equiv e\textbf{B}/(m_e\omega_{pe}c)$; $t\equiv\omega_{pe}t$ have been used. The radiated electric field vectors can be calculated using Eq.(\ref{erad_xt}) and Eq.(\ref{track_sa}) for a given field configuration and initial position and velocity of the particles.

\begin{figure}
	\centering
	\noindent\includegraphics[width=3.0in]{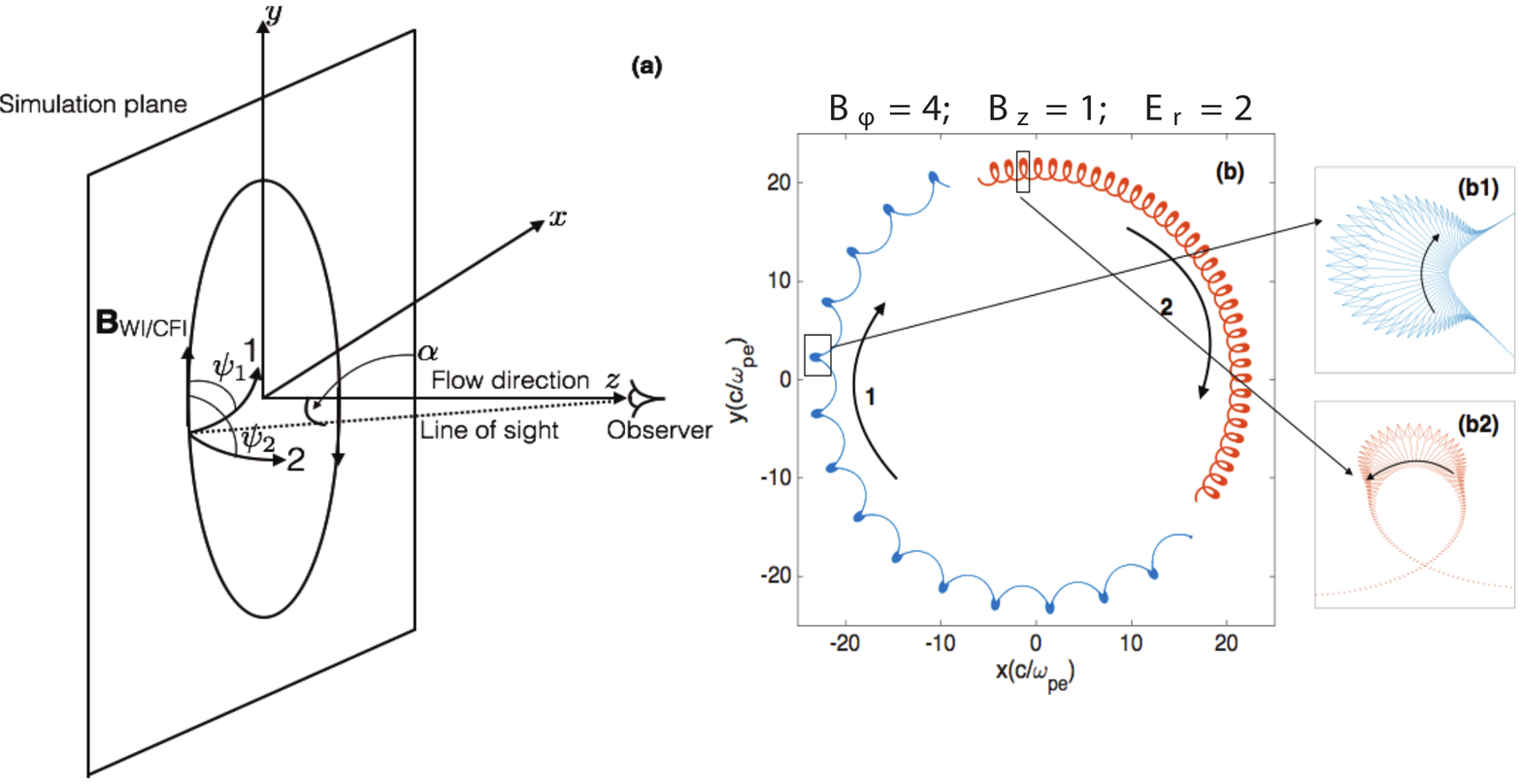}
	\caption{\label{fig2}(a) shows the schematic representation of the azimuthal magnetic field due to a cylindrical current filament with a flow along z with 1 and 2 representing electron trajectories that make the pitch angles $\psi_1<\pi/2$ and $\psi_2>\pi/2$ respectively with the direction of the WI/CFI magnetic field vector at the time of emission. (b) shows the trajectories of two electrons from the semi-analytic model for an observer along $\hat{\textbf{e}}_z$ (trajectory 1 ($\psi_1$) in blue and trajectory 2 ($\psi_2$) in red). The electric field vectors (arrows) are attached to each point on the trajectory, and shown more clearly in the zoom (b1) and (b2). It is clear that the radiation from trajectory 1 is left circularly polarized and from trajectory 2 is right circularly polarized because the electric field vectors are rotating clockwise and counter-clockwise respectively.}
\end{figure}

%(b1) and (b2) show the radiated electric field vectors obtained using the semi-analytic model for an observer along $\hat{\textbf{e}}_z$ for the electron making a pitch angle $\psi_1<\pi/2$ (trajectory 1) and $\psi_2>\pi/2$ (trajectory 2) in (b). Fig.b1 and b2 shows a section of the electric field vectors from trajectories 1 and 2 respectively.

A schematic representation of the geometry of the magnetic field due to the current filaments and the motion of plasma electrons in the frame of the background plasma is shown in Fig.~\ref{fig2}a. The radiated electric field vectors obtained using Eq.(\ref{erad_xt}) and Eq.(\ref{track_sa}) for electrons making pitch angles $\psi_1(<\pi/2)$ and $\psi_2(>\pi/2)$ with respect to the direction of WI/CFI magnetic field at the instant of emission is shown in Fig.~\ref{fig2}b. The direction of rotation of the radiated electric field vectors show a left handed (LH) circular polarization for $\psi_1$ and right handed (RH) for $\psi_2$. The handedness of circular polarization depends on the pitch angle. The emission from a single electron whose velocity vector makes a pitch angle $\psi$ with the direction of the WI/CFI magnetic field vector at the instant of emission is equivalent to that emitted by a particle moving at a constant speed in a circular path. For the observer lying on the z-axis in the far field, it is reasonable to assume that the angle between the observer and the point of emission $\alpha\approx$ 0. Hence,  the electric field vectors of the emitted radiation will be oriented in the xy plane. For a relativistic electron, the radiation lasts for a very short time and is limited within a small angle $\theta\sim1/\gamma_e\ll 1$ for a fixed observer. Under this assumption, Eq.(\ref{erad_omega}) reduces to,  $\textbf{E}_\mathrm{rad}=-2(i(1+\gamma_e^2(\pi/2-\psi)^2)/(\sqrt{3}\gamma_e^2)K_{2/3}(\xi)\hat{\textbf{e}_\mathrm{x}}+(\pi/2-\psi) \sqrt{1+\gamma_e^2(\pi/2-\psi)^2}/(\sqrt{3}\gamma_e)K_{1/3}(\xi)\hat{\textbf{e}_\mathrm{y}})$, where $K_{2/3}$ and $K_{1/3}$ are the modified Bessel functions,  $\xi=\omega\omega_0(1+\gamma_e^2(\pi/2-\psi)^2)^{3/2}/3\gamma_e^3$, $\omega_0$ is the gyrofrequency and $\omega$ is the frequency of radiation~\cite{jackson}. It is clear that the polarization of radiation is right-handed (RH) or left-handed (LH) according to $\psi\gtrless\pi/2$ and is consistent with the results from the semi-analytical model shown in Figs.2b1 and 2b2. For a plasma with no initial magnetic field component along $\hat{\textbf{e}}_z$, the electrons have a symmetric pitch angle distribution for $\psi\gtrless\pi/2$. Hence, the contribution to the RH and LH polarizations are equal, resulting in a net zero circular polarization. In a magnetized plasma, the initial magnetic field ($B_z$) along $\hat{\textbf{e}}_z$ bends the electron trajectory in the xy plane creating an anisotropy in pitch angle distribution. When $B_z$ is along positive $\hat{\textbf{e}}_z$, the number of electrons with negative pitch angles exceed those with positive pitch angles resulting in a net RH circular polarization and vice versa when the direction of $B_z$ is reversed.

\begin{figure}
	\centering
	\noindent\includegraphics[width=3.0in]{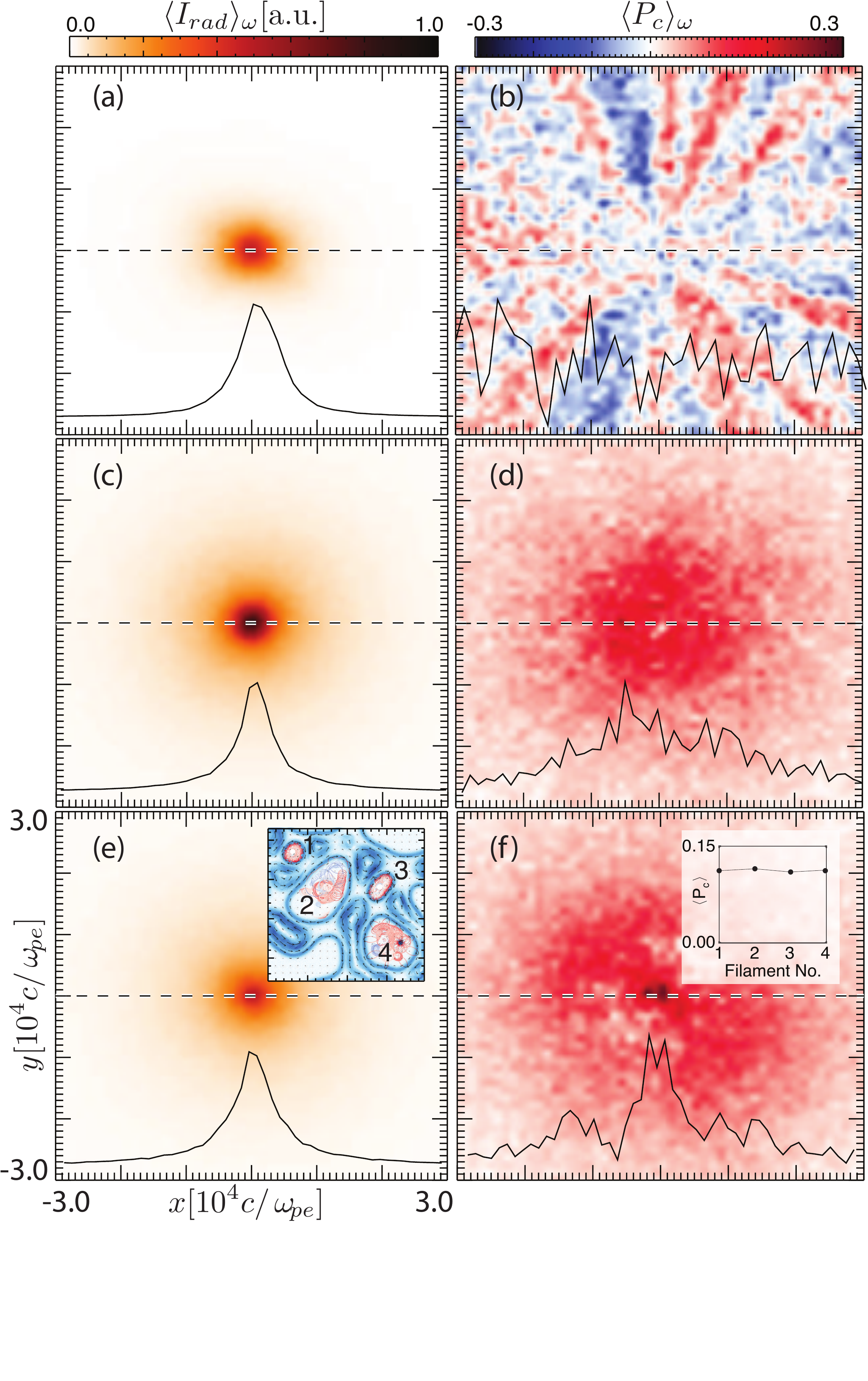}
	\caption{\label{fig3}Simulation results from jRad illustrating the key properties of radiation and degree of circular polarization from 1000 electrons trapped in a current filament. (a) and (c) show the frequency averaged spatial distribution of radiated energy on the detector lying in the $XY$ plane (parallel to the simulation plane) from the current filament shown in Fig.~\ref{fig1}c and 1b respectively. (b) and (d) show the frequency averaged degree of circular polarization corresponding to (a) and (c). For the unmagnetized filament, there is almost equal contribution to left and right circular polarization, whereas, a magnetized filament produces a strong right circularly polarized radiation with peak value $\sim 25\%$. (e) shows the frequency averaged spatial distribution of radiated energy on a detector similar to (c) from four magnetized current filaments ( with initial magnetization $\sigma$=0.05) of electron-proton plasma shown in the inset (a zoom of this inset can be found in the Appendix~Fig.~\ref{supp1}). (f) shows the frequency averaged degree of circular polarization corresponding to (e) with the inset showing the $\big<P_c\big>$ for each of the corresponding filaments.}
\end{figure}
To validate our model, we extracted trajectories of 1000 electrons from a current filament directly from PIC simulations and computed the radiation spectrum and the degree of circular polarization ($P_c$) of the radiation emitted from them using the post-processing radiation code jRad~\cite{jrad}. Although the OSIRIS simulation is 2D, we reconstruct the 3D trajectories as $z=\int p_z(t)/\gamma_e(t)dt$. The spectrum was calculated on a two-dimensional virtual detector in the xy plane at a distance $z=1.5\times 10^4c/\omega_{pe}$ and size $60000\times 60000(c/\omega_{pe})^2$ divided in $100\times 100$ cells to capture the entire emitting region. The detector captured a spectra of frequencies ($\omega$) in the range $\omega =(10^0-10^3)\omega_{pe}^{-1}$ with a resolution of 256 cells per decade in the frequency axis. Radiation was calculated following the trajectories from time $t_i=6300\omega_{pe}^{-1}$ to $t_f=9000\omega_{pe}^{-1}$ during which the filament was in a steady state. The degree of circular polarization ($P_c$) was estimated using the relevant Stokes parameters, in which $P_c=V/I$, $V=2\langle Im[(\boldsymbol{\epsilon}_1\cdot \textbf{E}_{rad})^*(\boldsymbol{\epsilon}_2\cdot \textbf{E}_{rad})]\rangle$, $I=|\boldsymbol{\epsilon}_1\cdot \textbf{E}_{rad}|^2+|\boldsymbol{\epsilon}_2\cdot \textbf{E}_{rad}|^2$ and $\boldsymbol{\epsilon}_1$ and $\boldsymbol{\epsilon}_2$ are the unit vectors perpendicular to the direction of observation. The angular brackets $\langle\cdot\rangle$ represent the time average. Figs.3a and 3c show the spatial distribution of frequency averaged radiated energy, $\langle I_{rad}\rangle_\omega=\int I_{rad}d\omega/\int d\omega$ from the electrons confined in the current filament of Figs.1b and 1c respectively. Figs. 3b and 3d show the corresponding $\langle P_c\rangle_\omega=\int P_cI_{rad}d\omega/\int I_{rad}d\omega$ for the filament. A strong right circular polarization with a peak value of $\langle P_c\rangle_\omega\approx 0.25 (25\%)$ is observed for the magnetized filament. For the unmagnetized filament, a nearly equal distribution of left and right circular polarization is observed resulting in a net zero circularly polarized radiation flux. The total circularly polarized radiation flux, $\langle P_c\rangle$, can be obtained using the formula $\langle P_c\rangle=\iint P_cI_{rad}dAd\omega/\iint I_{rad}dAd\omega$, where $A$ is the detector area. For the radiation in Fig.~\ref{fig1}b, $\langle P_c~\rangle=0.117 (11.7\%)$. To confirm that $\langle P_c\rangle$ was not affected by the time period on which it is averaged, the time of averaging ($6300\omega_{pe}^{-1}-9000\omega_{pe}^{-1}$) was divided into three equal segments of interval $900\omega_{pe}^{-1}$ and $\langle P_c\rangle$ was separately calculated for each of these intervals. We found that $\langle P_c\rangle$ was equal in all the cases with a variation of $\pm 0.1\%$, which indicates that our observation for a single filament are physically meaningful.

Furthermore, to check for the effect of multiple filaments, we computed the radiation spectra and $P_c$ from 1000 electrons distributed equally in four filaments of a magnetized electron-proton plasma of Fig.~\ref{fig1}b. Figs.~\ref{fig3}e and ~\ref{fig3}f show the $\langle I_{rad}\rangle_\omega$ and $\langle P_c\rangle_\omega$ respectively for the four filaments described in the inset of Fig.~\ref{fig3}e. The radiation spectra and circular polarization is similar to the single filament. In addition, the $\langle P_c\rangle$ obtained from each of the filaments (inset of Fig.~\ref{fig3}f) is computed separately and found to be nearly equal ($\sim 0.115$). This shows that the anisotropy in pitch angle distribution depends only on the fields and not the filament shape. An important observation is that WI/CFI in magnetized plasmas generate local islands of anisotropic pitch angle distribution. Hence, circular polarization can be observed even in case of WI/CFI driven shocks in magnetized spherical jets irrespective of the position of observation in contrast to the model described earlier~\cite{nava}, which is valid only for collimated jets.

\begin{figure}
	\centering
	\noindent\includegraphics[width=3.0in]{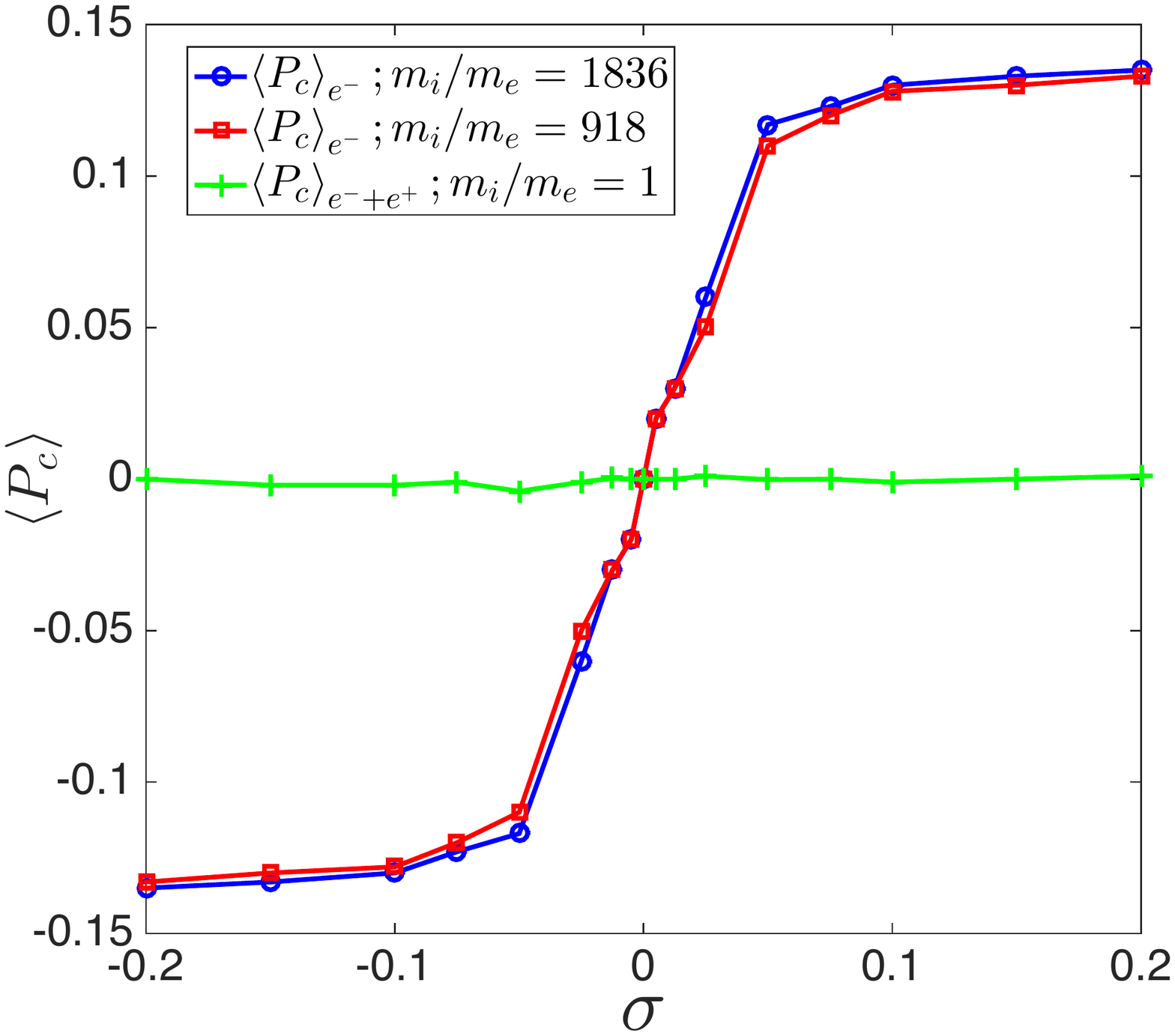}
	\caption{\label{fig4}Simulation results from jRad illustrating the normalized flux of circularly polarized photons (averaged over frequency and spatial domain) from 1000 electrons trapped in a typical current filament arising from interpenetrating flows of electron-positron and electron-ion plasmas for varying magnetizations. Note that here negative $\sigma$ represents negative $B_z$.}
\end{figure}
To investigate the role of initial magnetization of the plasma, $\langle P_c\rangle$ was calculated from 1000 electrons confined in a single filament of electron-proton plasma with initial magnetizations varying from 0.0 to 0.2 for both signs of $B_z$. It was found that the $\langle P_c\rangle$ increased initially with magnetization and then started to saturate beyond $\sigma = 0.05$ converging to $\sim 13\%$. %
The level of circular polarization depends on the electron pitch angle distribution, and we have confirmed that the external magnetic field changes this distribution such that finite levels of circular polarization become possible. Specifically, in unmagnetized scenarios, the pitch angle distribution is symmetrical about $\pi/2$. Thus, the total level of circular polarization vanishes in this case. In magnetized scenarios, instead, the pitch angle distribution is off-set due to the particle drifts that appear in the presence of the external magnetic fields. As the pitch angle distribution becomes asymmetric, the level of circular polarization level increases. %%
Other processes can also lead to finite circular polarization levels, for instance \cite{sinha} where generation of circular polarization in a different physical configuration is attributed to an assymetric energy dissipation mechanism instead of the topological changes in pitch-angle distribution discussed here. 
%This indicates that the electron velocity vector in the azimuthal direction competes with the velocity vector due to the cyclotron motion induced by the initial magnetization. The influence of cyclotron motion due to initial magnetizaton on the pitch angle increases with $\sigma$ leading to an increase in the degree of circular polarization.

For high magnetizations ($\gtrapprox 0.05$), the cyclotron motion completely dominates over the azimuthal velocity vector leading to a saturation of the pitch angle distribution which results in the saturation of the $\langle P_c\rangle$. The handedness of circular polarization changes with the sign of $B_z$ because the velocity vector due to cyclotron motion reverses with the sign of $B_z$ resulting in an anisotropy in the opposite direction. Furthermore, to understand the effect of plasma composition i.e. ion-electron mass ratios ($m_i/m_e$), we performed simulations for the same magnetizations for plasmas with $m_i/m_e=1$ and 918. We observed that the values of $\langle P_c\rangle$ for $m_i/m_e=918$ and 1836 were almost equal. This is because the ion response time scales are very large when compared to electron time scales, resulting in the radiation primarily emitted only by the electrons and the circular polarization arising only due to the anisotropies in electron pitch angle distribution. For electron-positron plasmas, both species contribute to the emitted radiation. They rotate in opposite directions due to their opposite charge and create anisotropies in mutually opposite directions resulting in a net cancellation of the circular polarization.

In conclusion, we have shown that the motion of the charged particles in the fields due to WI/CFI produces strong radiation emission in the direction of the plasma flow. An initial magnetization breaks the symmetry in pitch angle distribution of the electrons resulting in a partially circularly polarized radiation. Results indicate that the anisotropies in pitch angle distribution saturates at high initial magnetizations leading to a saturation in the degree of circular polarization. The emission of circularly polarized radiation is limited to electron-ion plasmas (plasmas with $m_i/m_e > 1$). For pair plasmas, the electrons and positrons produce circular polarization of opposite handedness resulting in a net zero circular polarization. The simulation set-up mimics the flow believed to be present in astrophysical scenarios. Hence, this study is significant to understand the origin of circular polarization in the recent observation of GRB afterglow~\cite{wiersema2014}. As the anisotropy in pitch angles is local to the current filaments, the treatment is valid even for broader or spherical jets. With the observation of WI/CFI in laboratory and future proposals to study the interaction of electrically neutral electron-positron beams (fireball beams) with plasma~\cite{huntington,fox2013,sarri}, it may be possible to design scaled experiments to test the results presented in this Letter and develop a better understanding of radiation and its polarization in real astrophysical scenarios. %We finally note that there are other configurations that allow for the production of circularly polarized radiation in WI/CFI scenarios based on a different physical mechanism (Sinha et al. to be submitted), thus stressing that the WI/CFI may play an important role in explaining the origin of circularly polarized radiation recently observed~\cite{wiersema2014}.

\acknowledgments
The authors would like to thank Prof. Mikhail Medvedev for very insightful discussions. This work was supported by the European Research Council (ERC-2010-AdG grant 267841 and ERC-2015 grant 695008) and FCT (Portugal) grant no. PTDC-FIS-PLA-2940-2014. J.V. acknowledges the support of FCT (Portugal) grant no. SFRH/IF/01635/2015. We acknowledge PRACE for awarding access to resource SuperMUC (Leibniz research center) and Fermi (CINECA). We also acknowledge the supercomputing resource IST Cluster at IST. 

\software{OSIRIS \citep{osiris,fonseca2013}, jRad \citep{jrad}}

\vspace{5mm}
%\facilities{HST(STIS), Swift(XRT and UVOT), AAVSO, CTIO:1.3m,


%\software{astropy \citepp{2013A&A...558A..33A},  
%          Cloudy \citepp{2013RMxAA..49..137F}, 
%          SExtractor \citepp{1996A&AS..117..393B}
%          }



\appendix
%\counterwithin{figure}{section}
\section*{Degree of circular polarization from multiple filaments}
\begin{figure}
	\centering
	\noindent\includegraphics[width=3.0in]{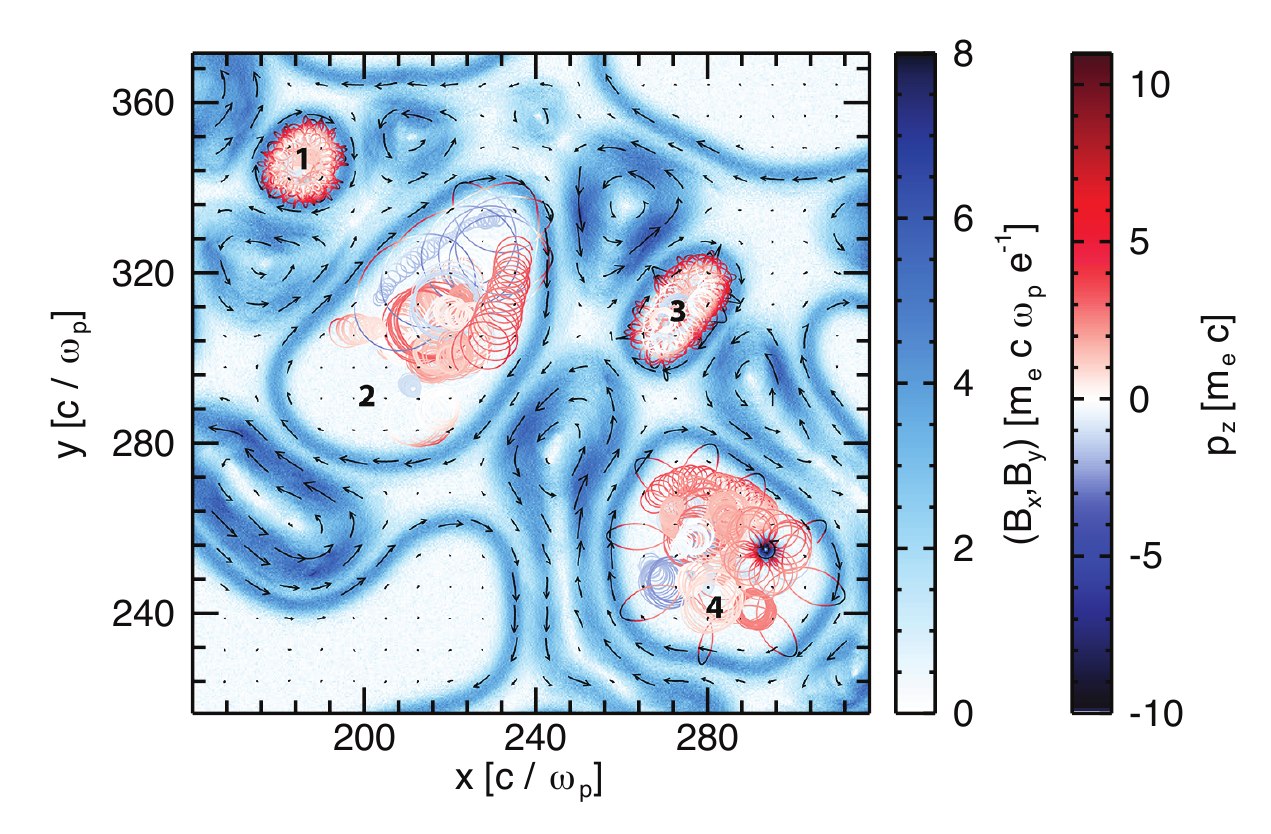}
	\caption{\label{supp1}The transverse magnetic field vectors (represented by arrows) arising due to WI/CFI in a magnetized interpenetrating electron-proton ($m_i/m_e=1836$) plasma flow with initial magnetization $\sigma=0.05$ at time t=9000$\omega_{pe}^{-1}$ is shown. The trajectories of 25 electrons each trapped in four different current filaments are shown from time $t_i$=8910$\omega_{pe}^{-1}$ to  $t_f$=9000$\omega_{pe}^{-1}$, with color scales representing their longitudinal momentum ($\textbf{p}_z$).}
\end{figure}

%\subsection{}
Although we have only considered radiation from a single filament, it is important to consider the net effect on the degree of circular polarization due to radiation from multiple filaments. A simple analytic model can be constructed by assuming the current filaments as independent sources of radiation. In the far field, each filament can be described as a point source of radiation with electric field components given by $E_x(r,t)=A\exp [i(\textbf{k}\cdot\textbf{r}-\omega t)]$ and $E_y(r,t)=A\exp [i(\textbf{k}\cdot\textbf{r}-\omega t+\phi)]$ with $\textbf{k}$ being the wave vector, $\omega$ the frequency and $\phi$ the phase difference between the components. The degree of circular polarization associated with a single filament is then, $P_c=\sin(\phi)$ where the light is circularly (elliptically) polarized for $\phi=\pi/2(\phi<\pi/2)$. To incorporate the effect of multiple current filaments, we consider the superposition of $N$ plane waves with a random phase factor $\psi_k$ between them. The resultant electric field components can be written as $E_x(r,t)=\sum_{n=1}^{N}A_{n}\exp [i(\textbf{k}_{n}\cdot\textbf{r}-\omega_{n}t+\psi_{n})]$ and $E_y(r,t)=\sum_{m=1}^{N}A_{m}\exp [i(\textbf{k}_m\cdot\textbf{r}-\omega_{m}t+\psi_{m}+\phi_m)]$. The degree of circular polarization ($P_c$) is given by the relevant Stokes parameters, in which $P_c=V/I$, $V=2\langle \mathrm{Im}\{E_x^*E_y\} \rangle$, $I=E_x^* E_x + E_y^* E_y$, and where the angular brackets $\langle\cdot\rangle$ represent the time average. Hence, $V = 2 \langle \textrm{Im} \{ \sum_{n,m=1}^{N} A_n A_m \exp \left[i \left(\Delta \mathbf{k}_{n,m}\cdot\textbf{r}-\Delta \omega_{n,m} t+\Delta \psi_{n,m}+\phi_m\right)\right]\}\rangle $, $I=\sum_{n,m=1}^N A_n A_m \exp \left[i\left (\Delta \mathbf{k}_{n,m}\cdot\textbf{r}-\Delta \omega_{n,m} t+\Delta \psi_{n,m}\right]\left[1+\exp(i \Delta \phi_{m,n}\right)\right]$, and $\Delta a_{n,m} = a_m-a_n$ where $a$ is a generic quantity. If the phase $\psi_n$ is randomly distributed, we can employ the random phase approximation~\cite{rpa}, for which $\langle\exp[i \Delta \psi_{n,m}]\rangle_{s}=\delta_{m,n}$, where $\langle\cdot\rangle_s$ is the average over a statistical ensemble of systems differing from one another only in the phase $\psi$. It is thus straightforward to show that $V=2\langle \sum_{n=1}^N A_n^* A_n \sum_{m=1}^N \sin(\phi_m)\rangle =\sum_{n=1}^N A_n^* A_n \sum_{m=1}^N \sin(\phi_m)$ and $I = 2 N \sum_{n=1}^N A_n^* A_n $. Hence, the degree of circular polarization emitted by $N$ filaments corresponds to:
\begin{equation}
P_c=\frac{1}{N}\sum_{m=1}^{N}\sin(\phi_m)\label{rpa}
\end{equation}
Equation~(\ref{rpa}) is consistent with our simulation results, where we tracked 1000 electrons distributed equally in four different filaments of the magnetized electron-proton plasma described in Fig.~\ref{supp1}. The inset of Fig.~\ref{fig3}f shows the $\langle P_c\rangle$ for the four filaments separately, the average of which is 0.113. The $\langle P_c\rangle =0.115$ when calculated from all the four filaments together. We emphasize that Eq.~(\ref{rpa}) is valid for an arbitrary large number of filaments, such that our results hold in such scenarios. %This confirms the validity of Eq.(\ref{rpa}).

%% thebibliography produces citations in the text using \bibitem-\citep
%% to the KEY in the \citep commands (see the first section above).
%% place of the \citep commands.



\bibliography{circularpolarization}

\end{document}